\documentclass[prb,twocolumn,showpacs,superscriptaddress,preprintnumbers,amsmath,amssymb]{revtex4-1}
\usepackage{graphicx}
\usepackage{dcolumn}
\usepackage{bm}
\usepackage{amsmath}
\usepackage{color}
\usepackage{hyperref}
\usepackage[utf8]{inputenc}
\usepackage[T1]{fontenc}

\begin{document}


\title{Interplay of thermal and quantum spin fluctuations
on the kagome lattice}

\author{Dirk Wulferding}
\author{Peter Lemmens}
\author{Patric Scheib}
\affiliation{Institute for Condensed Matter Physics, Technical University of Braunschweig, D-38106~Braunschweig, Germany}

\author{Jens R\"{o}der}
\affiliation{Institute for Physical and Theoretical Chemistry, Technical
University of Braunschweig, D-38106~Braunschweig, Germany}

\author{Philippe Mendels}
\affiliation{Laboratoire de Physique des Solides, Universit\'{e} Paris-Sud 11, UMR CNRS 8502, 91405 Orsay, France}

\author{Mark A. de Vries}
\affiliation{Department of Physics and Astronomy, University of Leeds, Leeds, LS2 9JT UK}

\author{Shaoyan Chu}
\affiliation{Center for Materials Science and Engineering,
Massachusetts Institute of Technology, 77 Massachusetts Ave,
Cambridge MA 02139, USA}

\author{Tianheng Han}
\author{Young S. Lee}
\affiliation{Department of Condensed Matter Physics, Massachusetts Institute of Technology, 77 Massachusetts Ave, Cambridge MA 02139, USA}

\date{\today}

\begin{abstract}
We present a Raman spectroscopic investigation of the Herbertsmithite ZnCu$_3$(OH)$_6$Cl$_2$, the first realization of a Heisenberg $s=1/2$ antiferromagnet on a perfect kagome lattice. The magnetic excitation spectrum of this compound is dominated by two components, a high temperature quasi elastic signal and a low temperature, broad maximum. The latter has a linear low energy slope and extends to high energy. We have investigated the temperature dependence and symmetry properties of both signals. Our data agree with previous calculations and point to a spin liquid ground state.

\end{abstract}


\maketitle

\section{Introduction}
Strongly frustrated spin systems are in the focus of many recent works as the competition of exchange interactions leads to interesting ground states and a rich low energy excitation spectrum. Of these systems, the $s=1/2$ Heisenberg antiferromagnet on the kagome lattice has been of special interest, since it displays a unique quantum disordered and highly degenerate ground state, i.e. a spin liquid \cite{review}. A kagome lattice consists of a two-dimensional (2D) arrangement of corner-sharing triangles. In systems with large spin so called weather-vane modes, low energy collective rotations of spins on a hexagon, destabilize any order. For $s=1/2$ a quantum superposition of an infinitely dense spectrum of low energy singlet and triplet excitations are supposed to exist. Defects on the kagome lattice may lead to a gapless valence bond glass phase \cite{singh10} with enhanced singlet bonds opposite to the defects. Together with antisymmetric Dzyaloshinskii-Moriya (DM) interactions a quantum critical point is induced \cite{Rousochatzakis}. Despite these and other \cite{marston} theoretical investigations, the understanding of the ground state properties and the excitation spectrum of the quantum kagome system is not completely satisfactory.

Earlier experimental investigations of compounds that are close in realizing the kagome layer are hampered either by interlayer interactions, a non-uniform, more complex distribution of exchange coupling constants or spin anisotropies. The first reported material with 2D kagome layers was SrCr$_8$Ga$_4$O$_{19}$ \cite{ramirez} with $s=3/2$. It \textbf{shows} a spin-glass transition for temperatures below $T = 3.3$ K despite the much larger interaction strength. In the $s=3/2$ compound Y$_{0.5}$Ca$_{0.5}$BaCo$_4$O$_7$ \cite{schweika} short ranged chiral spin correlations develop at low temperature, while the $s=1/2$ compound RbCu$_3$SnF$_{12}$ \cite{morita} experiences a spin gap that separates the singlet groundstate from low energy triplet excitations.

Presently the compound ZnCu$_3$(OH)$_6$Cl$_2$ is the only well characterized realization of the $s=1/2$ Heisenberg antiferromagnet on a perfect kagome lattice \cite{shores,mendels-rev}. It exists in two structurally slightly different forms: a) Kapellasite \cite{colman} and b) Herbertsmithite. Small single crystals, however, are up to now only available as Herbertsmithite. In this compound, the Cu$^{2+}$ ions form the triangular kagome lattice in the crystallographic ab-layer and are coupled by oxygen atoms via superexchange pathways, leading to strong antiferromagnetic correlations of about \textbf{$J \approx 180$ K}. Despite this strong coupling, susceptibility measurements reveal no magnetic ordering down to 50 mK \cite{helton}.

The planes of these kagome structures are separated by intermediate
layers of Zn and Cl atoms and can therefore ideally be considered as
2 dimensional layers with negligible magnetic interactions along the
third dimension. However, Zn/Cu antisite disorder of at about 6$\%$
\cite{devriesdisorder} is believed to be present in previously
studied Herbertsmithite samples. Hence, the presence of a small
concentration of impurities may affect the spin correlations in the
kagome plane.

Muon spin rotation ($\mu$SR) measurements on the paratacamite Zn$_{0.5}$Cu$_{3.5}$(OH)$_6$Cl$_2$ showed a partly frozen ground state \cite{mendels}, which however disappears for higher Zn/Cu ratio. For the Herbertsmithite, i.e. a Zn to Cu ratio of 1/3, $\mu$SR did not reveal any ordering down to 50 mK. Electron spin resonance measurements revealed DM interactions of about $D = 0.08 \cdot J$ magnitude \cite{zorko}, which causes a ferromagnetic-like increase of susceptibility at low temperature. These and other experiments have been performed either on natural (mineral) crystals or powder samples \cite{mendels-rev}. Only recently, Herbertsmithite single crystals were successfully grown via hydrothermal synthesis \cite{lee}.


Our approach of investigating the excitations in the kagome lattice is using inelastic light scattering (Raman scattering). This is a unique method in the sense that it is sensitive to low energy topological, electronic, or magnetic excitations. Singlet modes ($s=0$) can be observed, which may otherwise only be probed in specific heat measurements. In particular, it has recently been suggested \cite{cepas} that studying the polarization dependence of the magnetic Raman scattering contribution should allow to distinguish between different possible ground states such as a valence bond crystal or a spin liquid. Of further interest is the spectral distribution of magnetic scattering as characteristic low energy slopes or sharp modes may be observed as signs of the topological character of the spin correlations \cite{ko,laeuchli}.


\section{Experimental}
This study deals with two different types of samples: (i) Natural crystallites from mineral sources, which show a dark green color and are slightly transparent. A selected crystallite from this source with the approximate volume of about 0.5 mm$^3$ was studied via Buerger precession method \cite{staudemann} to confirm the sample's single domain structure and to uncover a natural surface that is parallel to the crystallographic ab-plane, in which the kagome lattice is realized. Its stoichiometry, Zn$_{0.8}$Cu$_{3.2}$(OH)$_6$Cl$_2$, is slightly shifted from optimum. (ii) A hydrothermally synthesized single crystal with a size of about 1 x 0.2 x 0.2 mm$^3$, which shows a very regular habitus, was studied in addition. It appears in light green color and high transparency and has a stoichiometry very close to the perfect kagome, i.e. ZnCu$_3$(OH)$_6$Cl$_2$.

Raman spectroscopic studies were performed in quasi-backscattering geometry with a solid state laser ($\lambda = 532$ nm) and 1 mW laser power. Room temperature measurements were carried out under ambient conditions, while low-temperature measurements were done in an evacuated, closed-cycle cryostat. Investigations of the polarization dependence were performed by rotating the sample within the ab plane. In the $0^\circ$ orientation, the a axis was approximately perpendicular to both the polarization and the $k$ vector of the incident laser light. The spectra were collected via a triple spectrometer (Dilor-XY-500) by a liquid nitrogen cooled CCD (HORIBA Jobin Yvon, Spectrum One CCD-3000V). In a previous experiment powder samples of Zn$_x$Cu$_{4-x}$(OH)$_6$Cl$_2$ (with $x = 0.5, 0.85, 1.0, 1.1, 1.2$ and $1.4$) were investigated to estimate the stoichiometry of the single crystals by comparing their phonon frequencies\cite{deVries}. These samples, however, do not allow to successfully study the weaker magnetic or topological Raman scattering in the Herbertsmithite.

\section{Results}
Figure 1 a) shows Raman spectra of the natural and of the synthesized single crystal, measured at room temperature ($T = 295$ K) and in $xx$ polarization.
Rotating the samples within the crystallographic ab plane (with the axis of rotation along the $k$ vector of the incident light) allows to distinguish between modes of A$_{1g}$ and E$_g$ symmetry representation. Furthermore, the previously discussed rotational anisotropy of the magnetic modes can be probed.

\begin{figure}
\label{figure1} \centering
\includegraphics[width=8cm]{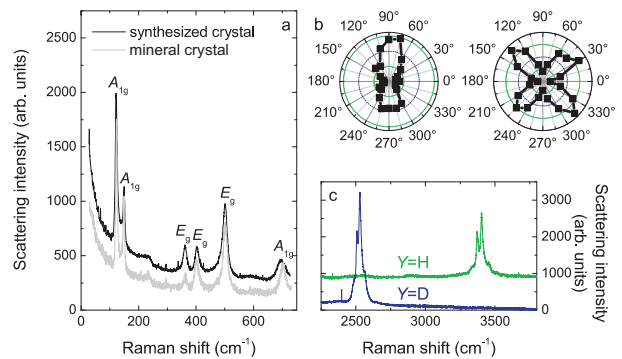}
\caption{a) Raman spectrum of natural and synthesized single crystal at room temperature in $xx$ polarization. b) left: intensity of the quasi elastic signal at RT as function of the angle between the light scattering polarization and a fixed direction within the ab plane as described in the text. The rotational anisotropy of this scattering intensity is consistent with that of A$_{1g}$ phonon modes; right: intensity of the E$_g$ mode at 363 cm$^{-1}$ plotted in a similar fashion. c) Comparison of phonon modes in ZnCu$_3$(O$Y$)$_6$Cl$_2$ assigned to H/D vibrations comparing a hydrated ($Y$=H) and a deuterated ($Y$=D) powder sample.}
\end{figure}

We observe 7 sharp modes at low to intermediate energies (123, 148, 365, 402, 501, 697-702 and 943 cm$^{-1}$) and five modes at high energies. We attribute these lorentzian lines to phonons. The factor group analysis yields 12 Raman active phonon modes according to the $R-3m$ space group: $\Gamma_{Raman}$ = 5 $\cdot$ A$_{1g}$ + 7 $\cdot$ E$_g$. In addition, we will later consider the A$_{2g}$ symmetry representations that can be induced by electronic resonances or DM interaction \cite{ko}. The corresponding tensors are:

\mbox{A$_{1g}$=$\begin{pmatrix}
a & 0 & 0\\
0 & a & 0\\
0 & 0 & b\\
\end{pmatrix}$

, E$_{g}$=$\begin{pmatrix}
c & 0 & 0\\
0 & -c & d\\
0 & d & 0\\
\end{pmatrix}$

, A$_{2g}$=$\begin{pmatrix}
0 & e & 0\\
-e & 0 & 0\\
0 & 0 & 0\\
\end{pmatrix}$.}

By rotating the samples within the ab plane ($x-y$ plane) the A$_{1g}$ modes are expected to give a rotation invariant contribution in parallel ($xx$) polarization and disappear in crossed ($xy$) polarization of incident and scattered light. The E$_g$ modes should show a four-fold symmetry in both polarizations. Due to the layered crystal structure and strongly anisotropic electronic polarizability perpendicular to the ab plane slight but inevitable misalignments of the crystals will show up as an additional two-fold modulation of the scattering intensity.


The 5 high energies phonons at about 3500 cm$^{-1}$ (2 $\cdot$ A$_{1g}$ + 3 $\cdot$ E$_g$) are due to vibrations of hydrogen atoms. In a deuterated powder sample they shift to lower energies with a renormalization factor of 0.74, which is in good accordance to a simple estimate ($\Delta \omega \approx \sqrt{m_H / m_D} = 0.71$). This is depicted in Fig. 1 c).

We assign the modes at 365, 402 and 501 cm$^{-1}$ to E$_g$ symmetry, as their intensity clearly displays a four-fold rotational symmetry (see Fig. 1 b). The phonon modes at 123, 148 and 702 (697) cm$^{-1}$ are distinctly different as they show a two-fold rotational symmetry. We assign these modes to A$_{1g}$ symmetry and the two-fold modulation of intensity to a small deviation of the $k$ vector from the c axis, i.e. a slight canting of the sample surface.

In a previous investigation of a series of Zn$_x$Cu$_{4-x}$(OH)$_6$Cl$_2$ powder samples we stated a linear dependence of the phonon mode at 702 (697) cm$^{-1}$ on composition. This allows us to estimate $x=0.8$ for the natural and $x=1.0$ for the synthesized single crystal, respectively. The synthesized single crystal is therefore in the regime of Herbertsmithite stoichiometry. Still, the natural single crystal is reasonably close to this state in the sense that thermodynamic and spectroscopic experiments did not observe evidence for long range ordering for $x=0.8$ \cite{mendels}. Studying the two different samples should therefore allow determining the influence of spin defects and static doping on the spin dynamics.

A less intense, asymmetric mode with a fano lineshape is observed at 230 cm$^{-1}$. This mode shows an even smaller intensity in the powder samples. Its lineshape may originate from the coupling of the corresponding phonon to a continuum of states. As its energy range is within the energy range of spin-fluctuations \cite{deVries2} and the compound is an insulator the corresponding continuum is attributed to magnetic fluctuations. The mode itself is probably induced by crystallographic disorder as the larger intensity modes already exhaust the total number of Raman active modes.

At low frequency and for high temperatures there is a pronounced quasi elastic scattering contribution (QES) visible with a two-fold rotational anisotropy (Fig. 1 b) that is similar to the phonon modes at 123, 148 and 702 (697) cm$^{-1}$, identified as A$_{1g}$ phonons. Therefore we assign the continuum to the same A$_{1g}$ component. At low temperatures (below $T = 50$ K) the quasi elastic scattering is completely depressed as demonstrated in Fig. 2 for the case of the natural sample.

\begin{figure}
\label{figure2} \centering
\includegraphics[width=8cm]{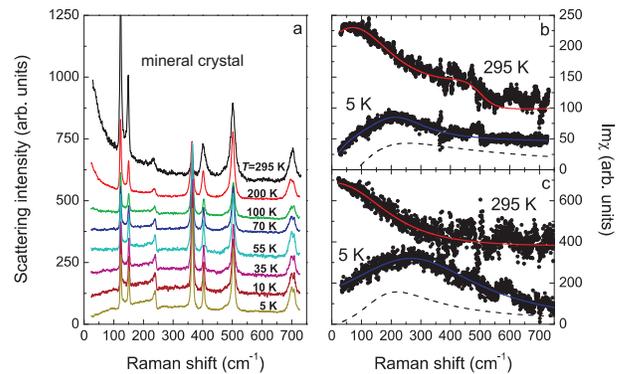}
\caption{a) Temperature dependence of Raman spectra of the natural sample in the frequency regime 30 to 725 cm$^{-1}$ and in parallel $xx$ polarization. Spectra are shifted for clarity. b) and c) show the Bose corrected spectra for the natural crystal and the synthesized sample, respectively, with phonon modes subtracted (dots) together with a fit to the background. The dashed black lines in b) and c) correspond to the fitted background in crossed $xy$ polarization at 5 K. The spectra at $T=295$ K in b) and c) are shifted in intensity for clarity.}
\end{figure}

\begin{figure}
\label{figure3} \centering
\includegraphics[width=8cm]{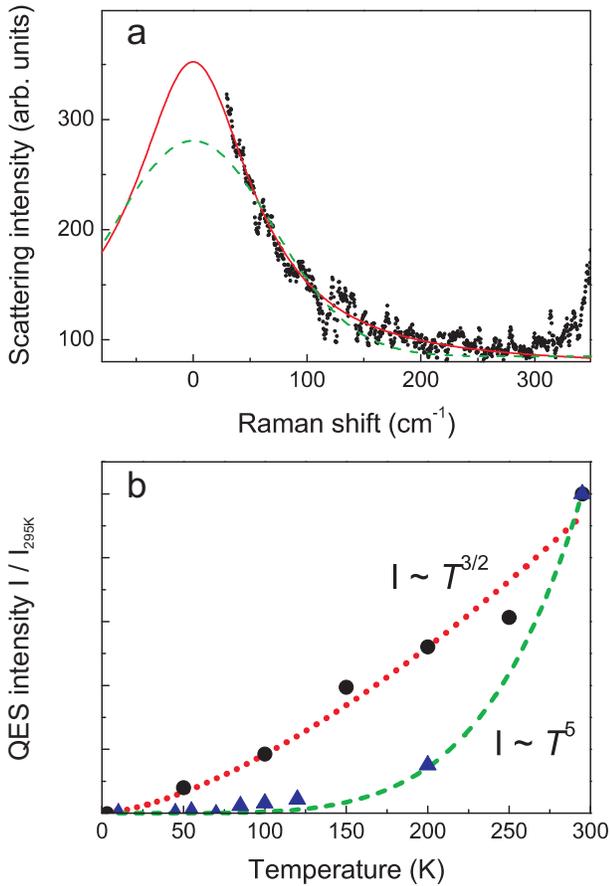}
\caption{a) Fit of Lorentzian (solid red line) vs. Gaussian (dashed green line) line shapes to the data (black dots) of the natural sample at low frequency and $T=295$ K. The phonon modes at 123 cm$^{-1}$ and 148 cm$^{-1}$ are subtracted. b) Temperature dependence of the intensity of the QES in the synthesized crystal (black circles) and the natural sample (blue triangles). The dotted (dashed) line is a fit to the data using a power law according to $I \sim T^{3/2}$ ($I \sim T^{5}$) for the synthesized (natural) sample as described in the text.}
\end{figure}

An analysis of the QES (black dots) at 295 K is given in Figure 3 a) together with a Lorentzian (solid red line) and a Gaussian (dashed green line) fit. The data is in better agreement with the Lorentzian line shape. Figure 3 b) shows the temperature development of the QES intensity in the synthesized (black dots) and the natural crystal (blue triangles). The decrease in intensity with decreasing temperature can be fitted with power laws (red and green curves). The observed exponent varies from $I(\omega \rightarrow 0,T$) $\sim$ $T^{3/2}$ to $I(\omega \rightarrow 0,T$) $\sim$ $T^{5}$ for the synthesized and the natural sample, respectively.

With decreasing temperatures we observe a moderate reduction of linewidth of all phonon modes, see Fig. 2 a. We attributed this to the decrease of anharmonic phonon scattering processes or other fluctuations. The integrated intensity of the asymmetric Fano line shape at 230 cm$^{-1}$ is increasing strongly; the intensity at 10 K is about 5x stronger compared to 120 K. In the synthesized sample this increase is only half that strong. At the same time, the linewidth of this mode is decreasing steadily before saturating for temperatures below approximately 70 K. With decreasing temperatures also the A$_{1g}$ phonons at 123 and 148 cm$^{-1}$ gain a Fano line shape.

At low temperatures a broad, finite energy maximum is observed with increasing scattering intensity. This continuum is observed for both the natural as well as the synthesized single crystal, as shown in Fig. 2 b) and c).
For the synthesized sample the maximum position is higher ($E_{max}=265$ cm$^{-1}\approx 2.1J$) compared to the natural one (215 cm$^{-1}\approx 1.7J$). In addition, the former sample also shows a more pronounced decrease of scattering intensity towards higher energies. We attribute the broad continuum in ZnCu$_3$(OH)$_6$Cl$_2$ to scattering on magnetic correlations as it resembles two-magnon scattering of a quantum antiferromagnet in the paramagnetic state \cite{lemmenspr}. Further below we will discuss and compare its properties in detail with existing theories. The dashed lines in Fig. 2 b) and c) show the corresponding results in crossed polarization. There are similar maxima and frequency shifts comparing the two sample qualities. However the onset of the scattering is shifted to higher energy for the natural sample.

We observe a larger spectral weight of the continuum in the synthesized sample, especially in $xx$ polarization. For both samples there is less spectral weight at low energies in $xy$ polarization. This suggests the possibility of decomposing the spectral weight into a low and a high energy part, to which the allowed symmetry channels A$_{1g}$, A$_{2g}$ and E$_g$ contribute to varying extend. To estimate their magnitudes we investigate the rotational anisotropies of these signals in Fig. 5. It is important to note that we observe no further signal that could be interpreted as of magnetic origin. In particular, there are no sharp modes at low energies. This statement is assured down to energy scales of approximately 5-10 cm$^{-1}$ as no upturn is visible at the edge of the experimentally accessible energy window.

\begin{figure}
\label{figure4} \centering
\includegraphics[width=8cm]{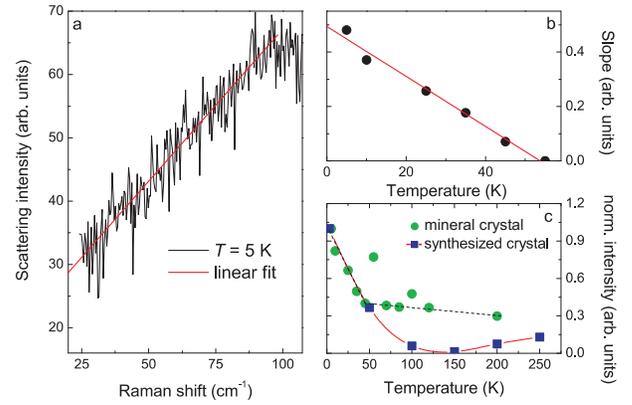}
\caption{a) Linear fit to the low frequency scattering in $xx$ polarization of the natural crystal at $T=5$ K. b) Temperature development of its slope including a linear fit. c) Temperature development of the intensity of the continuum in both the natural and synthesized sample with guides to the eye.}
\end{figure}

Figure 4 a) zooms into the low frequency regime (up to 100 cm$^{-1}$) at low temperature ($T = 5$ K). There is a linear frequency dependence towards the maximum and we can use its slope as a measurand of the integrated intensity, less sensitive to fluorescence backgrounds. In Fig. 4 b) the slope of the low energy scattering is plotted as function of temperature. It increases approximately linearly with decreasing temperature ($T < 50$ K). Both samples show this linear temperature dependence. However at higher temperatures the synthesized sample has again the more pronounced $T$ dependence as shown in Fig. 4 c). Such a linear increase of scattering intensity is anomalous and not compatible with the expected Bose factor. However, it is frequently observed for magnetic Raman scattering in low dimensional spin systems that are close to quantum critical.

\begin{figure}
\label{figure5} \centering
\includegraphics[width=8cm]{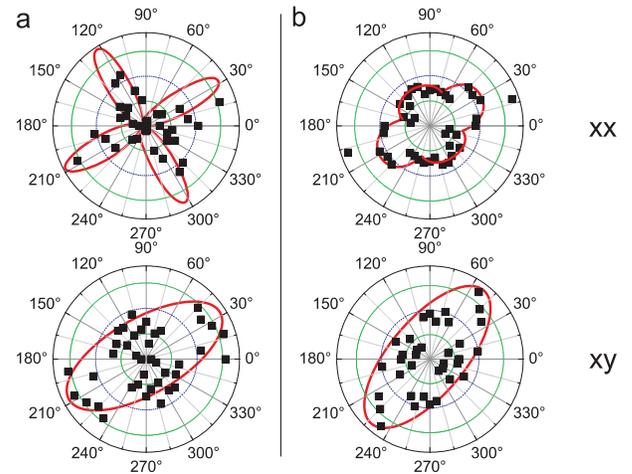}
\caption{Angular dependence at $T=5$ K of a) the low energy slope in the natural crystal and b) its continuum intensity. The data follow the rotational anisotropy expected for E$_g$ and A$_{1g}$ symmetry components. Note that measurements were performed only in the range from $0^\circ$ to $180^\circ$.}
\end{figure}

Finally, in Figure 5 we probe the rotational anisotropy of the low energy slope in $xx$ and $xy$ polarization (shown in a) and the rotational anisotropy of the intensity of the broad continuum (shown in b) in the natural sample at $T = 5$ K. The low energy slope in a) shows a four fold E$_g$-like symmetry in parallel polarization, while in crossed polarization the rotational anisotropy is rather two-fold or of mixed character. The intensity of the broad continuum in b) shows slight anisotropies in both $xx$ and $xy$ polarization.

\section{Discussion}

In the following we will first discuss the QES observed at high temperatures followed by the implications of the low temperature data at intermediate and high energies.

As mentioned above, the most characteristic feature of the QES is its strong depression of intensity with decreasing temperatures. Such a scattering is frequently observed in compounds that realize low dimensional spin systems, as spin chains \cite{kuroe} or frustrated dimer systems \cite{lemmenssrcubo}. It is related to energy density fluctuations that couple via spin-phonon coupling to the lattice and can be mapped on the specific heat $C_m$ of the spin system, $I(\omega \rightarrow 0,T$) $\sim$ $C_m$ $T^{2}$. Its lineshape and symmetry should be Lorentzian and of A$_{1g}$ symmetry, respectively \cite{lemmenspr}. The fit to the data given in Fig. 3 follows all these predictions very well. Again, the rotational invariance is distorted with two-fold symmetry but within experimental error identical to the one of the phonon modes identified as belonging to the A$_{1g}$ symmetry representation.

From the temperature dependence of the quasi elastic scattering intensity $I(\omega \rightarrow 0,T$) we can derive information on the evolution of low energy excitations. In, e.g., the Shastry-Sutherland compound SrCu$_2$(BO$_3$)$_2$ the exponential drop of the respective intensity has been used to derive a spin gap \cite{lemmenssrcubo}. The power law-like temperature dependence in ZnCu$_3$(OH)$_6$Cl$_2$ differs clearly from that and implies a spin liquid with more gradually evolving correlations \cite{hermele}. We attribute the softer power law $I(\omega \rightarrow 0,T$) $\sim$ $T^{3/2}$ in the synthesized single crystal to the reduced number of defects leading to an even more gradual evolution of spin correlations.

We will now discuss the three most recent theoretical investigations (using different approaches and techniques) of finite energy, magnetic Raman scattering in a kagome system and their implications. As was suggested by C\'{e}pas et al. \cite{cepas}, the formation of a valence bond crystal leads to a weakly broken translation symmetry of the spin system. Thereby the magnetic Raman response of the kagome plane should depend in a decisive way on the polarization of the incoming / scattered light, i.e. there should be characteristic oscillations of magnetic modes in the energy range of $J$ when the polarizations are rotated within the plane. In contrast, the ``true'' spin liquid exhibiting no broken symmetry should show a polarization independent magnetic Raman response. Since experimental investigations using different techniques have up to now not observed a spin gap, the Raman experiment is highlighted to be potentially sensitive to even weakly broken symmetries in kagome plane. It should be noted that in the second order perturbational treatment used by C\'{e}pas et al. magnetic scattering is only observed in E$_g$ symmetry, as the A$_{1g}$ component commutes with the used exchange scattering Hamiltonian.


Recently, calculations based on a modified Shastry-Shraiman model describing a $U$(1) Dirac spin liquid have been performed by Ko, et al. \cite{ko}. In this approximation also higher order terms, induced e.g. by electronic resonances of the phonon energy with correlated states of the system, are considered. Here, the included chiral $\vec{s} \times \vec{s}$ component leads to additional scattering contributions in A$_{1g}$, A$_{2g}$ and E$_g$ symmetry components. From the Raman tensors given above we notice that the now allowed A$_{2g}$ component is only observed in $xy$ polarization as a rotational invariant contribution. It does not contribute to $xx$ polarization at all. The spectral distribution of the Raman scattering cross section derived by Ko, et al. shows a broad continuum that extends up to about 600 cm$^{-1}$ ($\approx 5 \cdot J$).

Finally, L\"{a}uchli et al. \cite{laeuchli} calculated dynamical singlet fluctuations on the kagome lattice using large-scale exact diagonalizations on a 36 site lattice. The results show a broad continuum extending over a range of 2 -- 3$J$. There is also a pronounced intensity shooting up at lowest energies, $\omega / J \leq 0.2 \approx 25$ cm$^{-1}$, caused by a high density of low energy singlet and triplet excitations.



We can decompose the experimentally observed scattering continuum into two contributions both attributed to specific spin liquid excitations. The rotation dependence of the low energy slope (Fig. 5 a) shows in $xx$ polarization a clear E$_g$-like contribution, while in the $xy$ configuration it is oval shaped, most probably a mix of both E$_g$ and A$_{2g}$ symmetry. The dominance of the E$_g$ channel at frequencies below 100 cm$^{-1}$ is in contrast to theory \cite{ko}. The rotation dependence of the overall continuum's intensity (Fig. 5 b) shows a weak asymmetry in both $xx$ and $xy$ configurations for the natural and the synthesized sample. This anisotropy has no dominant four-fold contribution. However, a superposition of a four-fold and a two-fold contribution leads to a good description of the scattering intensity. We conclude that the A$_{1g}$ (for $xx$) and the A$_{2g}$ (for $xy$) symmetry channels are major contributions for the continuum at an energy of \textbf{$\approx 2J$}. These components are proposed to be due to higher order perturbation theory \cite{ko}. Comparing our experimental data in crossed polarization to the data in parallel polarization, no clear separation of the respective spectral weight is evident.

This behavior is in clear contrast to the phonon Raman scattering, where modes of A$_{1g}$ and E$_g$ show a pronounced selection rule and different in-plane anisotropy. The observed Fano-like line shape of phonons is also only observed for modes with A$_{1g}$ symmetry. On the other hand phonon and magnetic scattering differ within two respects that in the theoretical literature on the Herbertsmithite are jointly discussed \cite{Rousochatzakis}. One is related to defects, which for the local spin excitations play a different role than for \textbf{$q \approx 0$} optical phonons with large coherence length. The second aspect is DM interaction, which mixes singlet and triplet excitations. DM interaction and structural defects as intersite mixing will allow, further enhance and possibly mix A$_{1g}$ and A$_{2g}$ contributions, while spin defects in the plane reduce the scattering intensity. Therefore, a clear assignment of contributions from the A$_{1g}$, A$_{2g}$ and E$_g$ channels to the background is not straight forward. Still, the energy range and general shape of the observed background are in very good agreement with the above mentioned calculations of Ko, et al. \cite{ko}.

The effects of impurities in our Herbertsmithite samples should be
considered, and these effects should more pronounced for the natural
sample ($x=0.8$). While keeping the kagome plane intact, Cu atoms in
the intermediate layer may induce lattice deformations due to
Jahn-Teller distortion, as well as additional magnetic coupling to
the kagome Cu atoms. The effect of impurities in quantum spin
systems with dedicated low energy excitations can be exemplified
using the frustrated dimer system SrCu$_2$(BO$_3$)$_2$, where 1-2\%
of Zn doping on Cu broadens the low energy excitations and
considerably redistributes their spectral weight \cite{lemmenspr}.
This may be the reason that in contrast to theory, our experimental
data do not show sharp magnetic modes or a pronounced low energy
maximum.  In NMR experiments on Herbertsmithite it has been shown
that the local susceptibility in the proximity to a defect is
different and anomalous compared to the regular behavior
\cite{olariu}. With respect to Raman scattering it can be assumed
that the excitations in the continuum that consist mainly of
two-pair states \cite{ko}) should be reduced in intensity, comparing
the natural and the synthesized sample. Such a reduction of
continuum scattering intensity is indeed observed, as well as a
softening of about 20$\%$.


\section{Conclusion}

Our Raman data give further evidence that the algebraic spin liquid is the ground state of the Heisenberg $s=1/2$ kagome lattice antiferromagnet. This is based on the observation of a power law depletion of thermally induced fluctuations of the magnetic energy density with decreasing temperature and scattering continua in the low- and mid energy range. We give estimates of the contributions to these scattering continua to different symmetry components and conclude that except for the low energy continuum in $xx$ configuration, no strong E$_g$ contribution has been observed.

Our data leads to the conclusion that the kagome plane has short range and continuous correlations that form a maximum (at $\approx 2J$) with no clear cutoff but a moderate suppression at higher energy scales (at $\approx 6J$). The overall energy range and shape of the continua agrees well with existing calculations for a spin liquid state.

Presently the role of impurities in the single crystal, i.e. antisite disorder and local, defect-like interlayer interactions, is difficult to evaluate. Comparing the cleaner synthesized with the natural, natural samples a weak suppression of the continuum intensity with a frequency shift is evident. As no sharp, pronounced magnetic peaks are observed their relation to disorder remains an open issue.

\begin{acknowledgments}
This work was supported by DFG and the ESF program Highly Frustrated
Magnetism.  We like to thank W. Brenig, O. C\'{e}pas, P. Lee, C.
Lhuillier, E. Sherman, R. Singh and O. Tchernyshyov for important
and helpful discussions. D.W. acknowledges support from B-IGSM of
the TU-BS.  The work at MIT was supported by the Department of
Energy (DOE) under Grant No. DE-FG02-04ER46134.
\end{acknowledgments}

\end{document}